\documentclass[conference]{IEEEtran}
\usepackage{etoolbox}
\usepackage{placeins}
\makeatletter
\patchcmd{\@makecaption}
  {\scshape}
  {}
  {}
  {}
\makeatother
\IEEEoverridecommandlockouts
\usepackage{cite}
\usepackage{amsmath,amssymb,amsfonts}
\usepackage{algorithmic}
\usepackage{graphicx}
\usepackage{textcomp}
\usepackage{xcolor}
\def\BibTeX{{\rm B\kern-.05em{\sc i\kern-.025em b}\kern-.08em
    T\kern-.1667em\lower.7ex\hbox{E}\kern-.125emX}}
\begin{document}

\title{ReflectNet - A Generative Adversarial Method for Single Image Reflection Suppression\\
}

\author{\IEEEauthorblockN{Andreea Bîrhal\u{a}}
\IEEEauthorblockA{\textit{Adobe Systems Romania} \\
Bucharest, Romania \\
birhala@adobe.com}
\and
\IEEEauthorblockN{Ionu\c{t} Mironic\u{a}}
\IEEEauthorblockA{\textit{Adobe Systems Romania} \\
Bucharest, Romania \\
mironica@adobe.com}
}

\maketitle

\begin{abstract}
Taking pictures through glass windows almost always produces undesired reflections that degrade the quality of the photo. The ill-posed nature of the reflection removal problem reached the attention of many researchers for more than decades. The main challenge of this problem is the lack of real training data and the necessity of generating realistic synthetic data. In this paper, we proposed a single image reflection removal method based on context understanding modules and adversarial training to efficiently restore the transmission layer without reflection. We also propose a complex data generation model in order to create a large training set with various type of reflections. Our proposed reflection removal method outperforms state-of-the-art methods in terms of PSNR and SSIM on the SIR benchmark dataset.
\end{abstract}


\section{Introduction}
Capturing images through glass surfaces almost always produces images with unwanted reflections that ruin the content of the image. There are situations when no alternative exists to avoid this, for instance: when capturing city panoramas from the top floor of a tall building, the subject of the photo is protected by a case like in museums or when impressive landscapes are captured from a car or a plane during a travel. Such corruptions are difficult to retouch manually with existing image processing tools because reflections can completely cover regions from the image. Reflections not only affect image quality but it also modify the results of computer vision applications such as classification or segmentation. Thus, this challenging problem has attracted considerable attention from the computer vision community where their base approaches rely on Deep Convolutional Neural Networks~\cite{zhang2018single, bdn, errnet}. 
%The approach of \cite{zhang2018single} uses a fully convolutional network trained with perceptual losses to exploit both low-level and high-level information. \cite{bdn} propose a cascade deep neural network which estimates both the background image and the reflection layer. Method proposed in \cite{errnet} uses semantic clues to better discriminate whitin areas containing strong reflections.

The mathematical model of an image corrupted by reflection \textit{I} can be expressed as a linear combination of a transmission layer \textit{T} and a reflection layer \textit{R}:
\begin{equation}
I =  \alpha * T +  ( 1 - \alpha) * R \label{eq1}
\end{equation}
When given access only to $I$, is it obvious that there is an infinite number of possible decompositions.  Thus, the extraction of the background layer free of reflection artifacts is an ill-posed problem when a single image is given. Another challenge is the lack of real world training data because of the difficulty of obtaining ground truth labels.
Collecting real reflection images together with ground-truth transmitted layer is difficult because it requires  a mobile transparent surface and two consecutive data acquisitions: one for capturing the reflection image throw a glass plane and one for capturing the background after the glass has been removed. For obtaining paired training data for supervised learning, these acquisitions needs to capture exactly the same scene. But the scene can be changed even after small time differences because of: moving objects in the scene, lighting conditions, the characteristics of the glass surface that can distort the background. Thus, obtaining a large real-world dataset with aligned pairs is a tremendously labor intensive task. 
Therefore, previous learning-based approaches resort to the generation of synthetic images that model the physical formation of reflection images. However, existing image synthesis methods cannot perfectly model all the real cases, resulting in a discrepancy between the accuracy obtained by the learning methods on synthetic data and the accuracy obtained on real data.

In this paper we propose a novel single image reflection removal network designed to capture contextual information in order to achieve an efficient separation between reflection and transmission layer.  Our architecture is based on a ResNet~\cite{resnet} skeleton to which we added context modules to help network to learn a global representation of the input. We improved the classical ResNet architecture by including a channel attention mechanism at the end of each residual block. Therefore, although each convolutional filter has its own local receptive field, the output of the residual block also takes into account the global information. We further added dilated convolutions which support exponential expansion of the receptive field without loss of resolution. We optimize our network using a combination of four losses including pixel loss, perceptual loss, gradient loss and adversarial loss. When training our network, we applied different reflection models to generate as various synthetic reflection images as possible. Thus, the contributions of our paper are summarized below:
\begin{itemize}
\item We propose a novel network architecture designed to capture contextual information which we proved to be very useful in reflection removal task.
\item We propose a complex data generation method capable of producing realistic reflection scenarios. 
\item We use a complex objective function composed by a list of well known adapted losses.
\item We achieve better performance than state-of-the-art methods.
\end{itemize}

The rest of the paper is organized as follows. Section II describes related work in the reflection removal literature. Section III details our proposed approach, while Section IV presents experimental results and comparison with state-of-the-art methods. Finally, Section V concludes the paper.

\section{Related Work}
Single image reflection removal is a very ill-posed problem that has been studied during the last years. The first proposed algorithms rely on certain priors or additional information to reduce the complexity of the problem. Some methods simplify the problem by using multiple observations of the same scene as additional information to recover the transmission layer~\cite{multipleimages}.
Others use pairs of input images captured under different conditions e.g., flash/non-flash~\cite{flash} or captured through a polarizer at multiple orientations~\cite{polarization}. However, in many real scenarios, we do not have access to more images of the same scene and thus, these methods are not suitable for practical use. 

Other approaches use additional priors in order to simplify the problem of single image reflection removal. Method proposed in \cite{userassisted} utilizes a set of user annotations to guide layer separation together with a gradient sparsity prior. A Multi-Scale Depth of Field confidence map that guides the classification of edges is further used to reconstruct the background layer in~\cite{depthoffield}. The work of~\cite{ghostingcues} studies the ghosting cues which is a specific phenomenon when the glass is of  a certain thickness, in order to distinguish the reflection layer from the transmission layer.

Recently, there is an emerging interest in applying deep convolutional neural networks for solving single image reflection removal problem. Methods based on Generative Adversarial Networks (GANs) \cite{gans} have been recently applied in reflection removal task and obtained impressive results likewise in other image-to-image translations problems. In GANs, two networks are jointly trained in order to improve each other. The discriminator network is updated to distinguish real samples from the fake samples produced by generator network and generator is updated to generate data to fool the discriminator. Reflection removal methods based on GANs were applied to alleviate the problem of obtaining data pairs using unpaired data. Also, GANs-based methods are applied to produce more realistic results than other methods since adversarial loss encourages networks to generate images that look as if they came from natural image distribution.

Method proposed in \cite{zhang2018single} uses a deep neural network trained with a loss function composed of: feature loss, adversarial loss and exclusion loss. Feature loss is used to improve the visual perception of the network, adversarial loss refine the output transmission layer and the exclusion loss is designed to separate the background and transmission layer in the gradient domain.
In \cite{bdn}, proposed method not only predict the background layer but also the reflection layer.
The motivation of using a bidirectional estimation model is the mutual dependence of the estimation of reflection images and background images. They conclude that if a good estimation of the reflection image is provided, then it will be easier to estimate the
background image. They provides additional supervision signals to train the network by including the objective of recovering the reflection image.
The approach of \cite{errnet} is based on the observation that a semantic understanding is needed in order to separate the background layer from the reflection layer. They embrace context understanding modules that been effective used in semantic segmentation task. They also combat the problem of insufficient real training data pairs by proposing a framework for misaligned training data.

The work of~\cite{beyondliniarity} studied the existing data generation methods and propose a novel synthesis method using an additional neural network. They observed that the linear method for data generation used by the vast majority of computer vision community, presented in Equation~\eqref{eq1}, do not model the physical phenomenon of reflections, which is better modeled by a non-linear combination between background and reflection layers. Thus, they exploit GANs to create realistic and diverse training data. Method proposed in \cite{ghosting_cues} reduce the ill-posedness of the problem by focusing only on ghosting reflections. They use the ghosting cues that arise from shifted double reflections of the reflected scene off the glass surface to exploit the asymmetry between layers. To model the ghosted reflection a double-impulse convolution kernel is used. A different data generation model is also proposed in \cite{ceilnet}. They apply a Gaussian blur kernel with randomly piked $\sigma$ to handle a wide range of real cases. The brightness overflow from the naive image mixing is avoided by subtracting an adaptively computed value followed by clipping.

However, the vast majority of proposed reflection removal solutions are not able to completely restore the content of the images corrupted by reflections. Following a careful analysis of the existing methods and what they are missing in order to create a viable solution, we propose a series of innovative elements that complete the research in this field.
% \begin{itemize}
%     \item We propose a data generation model that covers this need, generating a wide range of real reflections. We observed that most of the existing methods generate synthetic reflection data that not model the phenomenon of reflection as it appears in reality and therefore they face performance problems when the model is tested on real data. We strongly believe that training data should simulate physical reflections as best as possible in order to create a solution usable for real scenarios.
%     \item We introduce context modules to help the network to reconstruct hidden or missing information. Existing GAN-based methods have a lack in understanding the global context of the input image. This proves to be a key element because we humans are able to separate transmission from reflection by understanding the content of at least one of the two layers.
%     \item We also concern about the size of the proposed model because our aim is to develop a solution for practical use. Thus, we carefully designed the model so that it outperforms state-of-the-art methods  using a limited number of learning parameters.
% \end{itemize}
We observed that most of the existing methods generate synthetic reflection data that not model the phenomenon of reflection as it appears in reality and therefore they face performance problems when the model is tested on real data. We believe that training data should simulate physical reflections as best as possible in order to create a solution usable for real scenarios. Thus, we propose a data generation model that covers this need, generating a wide range of real reflections. Another observation is that existing GAN-based methods have a lack in understanding the global context of the input image. This proves to be a key element because we humans are able to separate transmission from reflection by understanding the content of at least one of the two layers. Thus, we introduce context modules to help the network to reconstruct hidden or missing information. We also concern about the size of the proposed model because our aim is to develop a solution for practical use. Thus, we carefully designed the model so that it has impressive performance using a limited number of learning parameters.

% \textit{TODO: Aici as incerca sa pun 2 propozitii despre diferenta intre ceea ce facem noi si SOA: (1) majoritatea paper-urilor nu iau in considerare metode noi de generare a datelor. Noi am observat ca metode inteligente de generare a datelor si acordarea de metode suplimentare ajuta destul de mult. (2) desi multi folosesc GAN, majoritatea metodelor nu tin cont de context; (3) Aici a fost o agenda mai mult ascunsa a mea si am incercat sa te indrum fara sa iti dai seama (desi ai realizat de multe ori) este viteza si dimensiunea modelului (aici as vrea sa punem o mica sectiune in partea de SOA / sau dupa SOA pentru ca e relevant)}

\section{Approach}

Given an input image \textit{I} corrupted by reflection, our goal is to estimate a reflection free  transmission layer. To achieve this, we optimize a convolutional neural network $G_{\theta_G}$ with parameters $\theta_G$ to minimize the reflection removal loss function $L$.  Given training pairs ${(I_n, T_n), n = 1, N}$,  we aim to find the solution of:
\begin{equation}
\theta_G =  arg min_{\theta_G} \frac{1}{N} \sum_n L(G_{\theta_G}(I_n), T_n) \label{eq2}
\end{equation}
In the following, we will introduce our data generation model used to build training pairs, the proposed neural network architecture and the reflection removal loss function $L$.
\subsection{Data Generation}
The main challenge of the reflection removal problem is the insufficiency of a real large data set for training due to the complexity of the acquisition of paired data. The problem comes from the fact that collecting real reflection images together with ground-truth transmitted layer is difficult because it requires  two consecutive data acquisitions: one for capturing the reflection image throw a glass plane and one for the background after the glass has been removed. But the scene can be easily changed from one time to another because of natural phenomenons, like moving the camera or the scene objects. Zhang et. al~\cite{zhang2018single} proposed a benchmark dataset for reflection removal task composed by a limited amount of real paired data. Another dataset was released in \cite{SIR2-iccv17} with a little more data, but also for benchmarking purposes. Thus, there is not sufficient real paired data in order to update a neural network in a supervised manner.

In order to train our deep neural network, we generate a large dataset composed by reflection-transmission pairs. For image synthesis we used a non-linear method based on the following equation:

\begin{equation}
I =  \alpha \cdot T +  ( 1 - \alpha) \cdot ( K \otimes R ) \label{eq3}
\end{equation}

where $T$ is transmission layer, $R$ the reflection layer, $\alpha$ is a blending operator chosen in range $(0.3, 0.5)$, $K$ is a kernel that may have different forms depending on the thickness of the glass and $\otimes$ represents the convolution operator.

Reflections, depending on how they manifest in reality, can be mainly classified into 3 types: focused, defocused and ghosting as presented in \cite{beyondliniarity}.

\textbf{Focused reflection. }
This type of reflection appear when the object behind the glass and the reflected object are in the same focal plane. In this case, the reflection layer will be as sharp as the transmission layer in the reflection image.  For modeling this type of reflection, kernel $K$ is considered as one-pulse kernel. Thus, both transmission and reflection layers will look sharp and difficult to separate even by human eyes.

\textbf{Defocused reflection. }
Defocused reflection arrive when the image is captured from a certain distance to the camera, therefore the reflected objects and the subject objects are not in the same focal plane. In this case the reflection layer is out of focus and appear smoother than the transmission layer.  To simulate this type of reflection, kernel K is set as a Gaussian kernel.

\textbf{Ghosting reflection. }
When the thickness of the glass is not non-negligible, appear the refraction phenomenon and cause a quadratic reflection with shifting. To model this type of reflection we used for kernel $K$ a two-pulse kernel which is called ghosting kernel. 

\textbf{Saliency extraction. }
We observed that in most cases, the reflected object is the main object from the reflection layer. To highlight the main object we extract the saliency map S of the reflection layer which will be a $W_R \times H_R$ matrix with elements in interval $[0, 1]$ which will weight the contribution of the reflection layer at each pixel.
% (depending on the importance of the pixel on that position) 
Then, to generate the reflection image we use the following method:
\begin{equation}
I =  \alpha \cdot T +  ( 1 - \alpha) \cdot ( K \otimes (S \cdot R) ) \label{eq4}
\end{equation}
For saliency extraction, we used DeepLabV3 architecture \cite{deeplabv3}.

\textbf{Decrease saturation. }
We also observed that, in reality, the reflected objects are more desaturate than the objects from transmission layer. Thus, we decrease the saturation of reflection layer when generate synthetic data. To do this, we convert the image from RGB space to HSV space in order to have access to the saturation of pixels. We extract the saturation channel S, and decrease the values with a factor f that is chosen in range $[0.5, 0.8]$. Then, the image is converted back in the RGB space and used as a reflection layer. 

\begin{figure}
  \centering
  \includegraphics[width=0.45\textwidth, height=9cm]{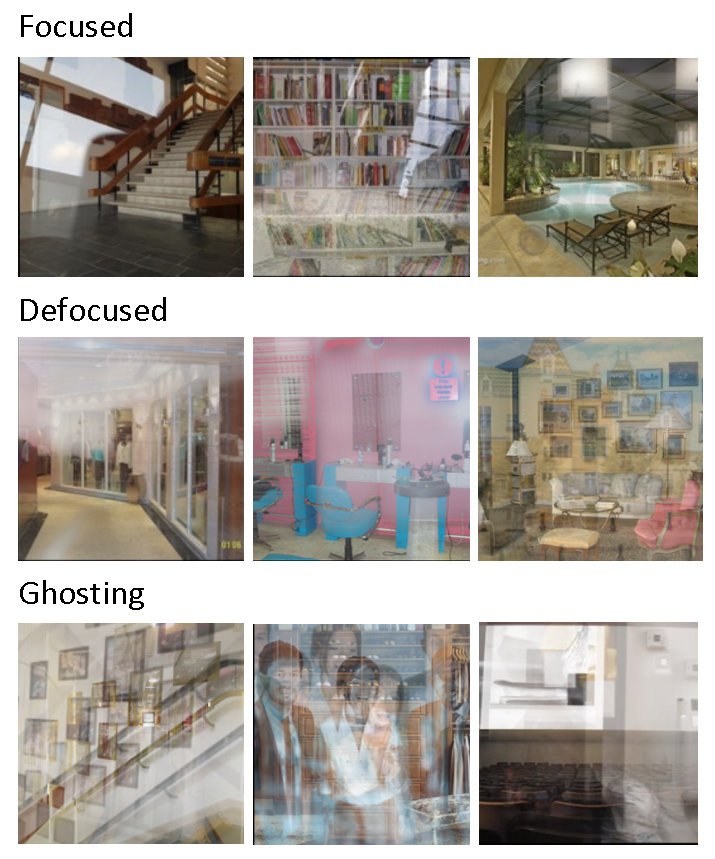}
  \caption{Examples of our generated data.}
  \label{fig:generated_data}
\end{figure}

\subsection{Architecture}

\begin{figure*}[ht!]
  \centering
  \includegraphics[width=\textwidth]{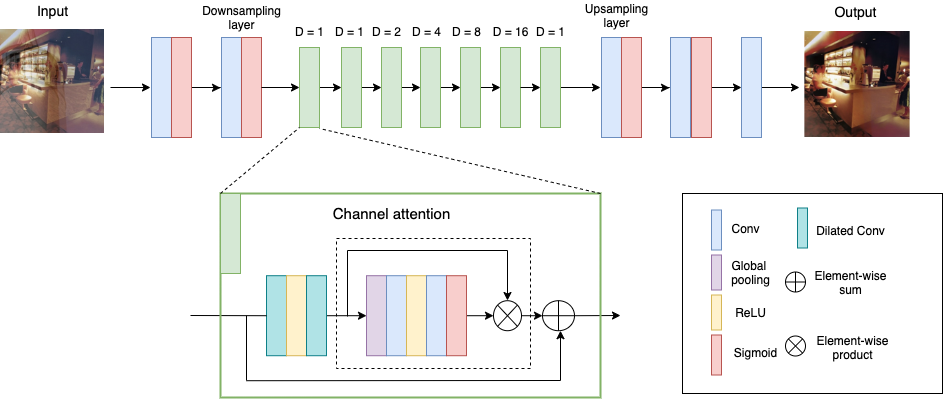}
  \caption{Proposed architecture.}
  \label{fig:architecture}
\end{figure*}

Human brain is able to reconstruct the information behind a reflection due to its ability to semantically understand the content of images. Inspired by how humans can mentally separate reflection layer from background layer by understanding the content of the two layers, we proposed a convolutional network architecture that is able to capture contextual information from the input image. Figure~\ref{fig:architecture}
illustrates the proposed architecture.

\textbf{Channel Attention module.}
Attention mechanism was first introduced in \cite{channelatt} and acts as a guidance to bias learning to focus on the most informative components of an input. 
Each convolution layer from a CNN operates with a local receptive field. Thus, the output after convolution is not able to exploit contextual information outside of the local region.  By adding channel attention mechanism, global spatial information is taken into account to determine the importance of each channel.  
Let $ X = [x_1, x_2, ..., x_C] $ be the activation produced by a network block, with $C$ feature maps with size of $W \times H$.  These activations reflect only the local information captured by corresponding receptive fields of each filter. To collect the global information, a global average pooling operator $f_{GP}$ is applied to each feature map $x_i$ and is obtained a scalar $z_i$:
$z_i = f_{GP}(x_i)$.
Channel specific descriptors $z_i$ compose the vector $z$ which will represent a statistical summary of global channel activations. Then we introduce a gating mechanism composed by a channel-downsampling trainable operator $W_D$ that will reduce the dimension of $z$ followed by a channel-upsampling trainable operator $W_U$ to revert to the original size. After downsampling, ReLU activation $\delta$ is applied, while after upsampling is applied a sigmoid activation $\sigma$. Thus, we compute channel-specific statistics $s = \sigma(W_U(\delta(W_D z))$, which we will use to calibrate each feature map via: $\hat x_i = s_i * u_i$.
We include this channel attention mechanism at the end of each residual block from our network architecture. Therefore, although each convolutional filter has a local receptive field, the output of a residual block also takes into account global information. We find capturing contextual information is very important to achieve an efficient separation between reflection and background layer.

\textbf{Dilated convolutions.}
Dilated convolutions are introduced in \cite{dilatedconv} as a necessity to aggregate multi-scale contextual information without losing resolution. They were successfully used in semantic segmentation task, where they have shown an impressive increase in performance over existing methods. The key of this innovation is that the number of parameters remain constant while the receptive field grows exponentially when the dilation size grows linearly. Inspired by their ability to capture contextual information, we introduced a context module based on dilated convolutions. The context module is achieved by progressive growing the receptive field of the convolutional layers using different dilation factors. Thus, our network contains in the middle $7$ residual blocks with the following dilation distances: $1, 1, 2, 4, 8, 16, 1$. We performed a series of experiments with different combinations of dilation blocks and we concluded that this configuration leads to best performances. We have found that the multi-scale context module helps the network to learn a more global representation of the input image so that an efficient separation between reflection and background layers is achieved.

\subsection{Losses}
The objective function is composed by four kinds of losses in order to separate transmission and reflection layer effectively:

\begin{equation}
L =  w_1  l_{pixel} + w_2  l_{perceptual}+ w_3  l_{gradient}  + w_4 l_{adv}  \label{eq6}
\end{equation}

\textbf{Pixel loss.}
We applied pixel loss to compare the pixel-wise difference between the output of the network $\hat T$ and the ground-truth transmission layer $T$. To ensure the similarity between the output and the ground-truth we use the Mean Square Error (MSE) to compute the pixel loss:
\begin{equation}
l_{pixel} =  || \hat T - T ||_2 \label{eq7}
\end{equation}

\textbf{Perceptual loss.}
We measure the perceptual distance between the output and transmission layer using a pretrained VGG-19 network \cite{vgg} to extract semantic features. We calculate the L1 loss between the feature vector of generated image and the feature vector of the ground-truth image.  Let $F_l$ be the feature from the $l$-th layer of VGG-19, we define the perceptual loss as:
\begin{equation}
l_{perceptual} =  \sum_l || F_l(\hat T) - F_l(T) ||_1 \label{eq8}
\end{equation}

where $l$ can be 'conv2\_2', 'conv3\_2', conv4\_2' and 'conv5\_2'  layers of VGG net.

\textbf{Gradient loss.}
The reflection image is the result of combining a reflection layer and a transmission layer. Thus, in most cases, an edge in the reflection image is generated either by transmission or reflection layer.  To achieve an effective separation, we include the gradient loss in our training that minimize the difference between gradients of generated image and gradients of ground-truth transmission. We compute both horizontal and vertical gradients and define the gradient loss as:
\begin{equation}
l_{gradient} =  || G_x(\hat T) - G_x(T) ||_1  + || G_y(\hat T) - G_y(T) ||_1\label{eq9}
\end{equation}

\textbf{Adversarial loss.}
To combat the insufficiency of the real paired data, we involve an adversarial loss in our training process. We collected a large dataset of reflection free images and used them as real samples for training the discriminator. In this way, the generator network G tries to generate realistic reflection free images, while discriminator network $D$ aims to distinguish between fake generated images and real-world reflection free images. We define an opponent discriminator $D$ as a VGG \cite{vgg} architecture while our proposed network acts the role of the generator. We minimize the relativistic adversarial loss proposed in \cite{relativistic_discriminator} due to the proved stability and generation of high quality data in GAN-based methods. Thus, the adversarial loss is defined as:
\begin{equation}
\!
\begin{aligned}
l_{adv}^G = -\mathbb{E}_{(x_r, x_f) \sim (P, Q)} [\log(\sigma(C(x_f) - C(x_r)))] \\
l_{adv}^D = -\mathbb{E}_{(x_r, x_f) \sim (P, Q)} [\log(\sigma(C(x_r) - C(x_f)))] 
\end{aligned}\label{eq:gan}
\end{equation}
where $P$ is the distribution of real data, $Q$ the distribution of fake data, $(x_r, x_f)$ is a pair sample from the two distributions, $\sigma(\cdot)$ is the sigmoid function and $C(\cdot)$ the non-transformed discriminator function as described in \cite{relativistic_discriminator}. 
\section{Experiments}

\subsection{Implementation details}
The use of real paired data in the training process has shown an increase in the performance of the reflection removal models. Since suficiently large real training data is not available in the research community, we adopt a fusion of synthetic and real data as our training dataset. We have collected a large amount of data with both indoor and outdoor scenarious and used them to generate synthetic images using our proposed generation model. We also include in our training set $90$ real-world training images from the dataset proposed in \cite{zhang2018single}. In total, our training set is composed by $29,000$ paired data.
Our implementation is based on Pytorch. We used Adam optimizer and train our model for $90$ epochs. The initial learning rate is set to $0.001$ and reduced over time.  We used Xavier method for weights initialization.

In the following experiments we used images from SIR benchmark dataset \cite{SIR2-iccv17} which contains both controlled and wild scenes. Controlled images are captured in a laboratory environment and contains subjects like postcards and solid objects, while wild scenes are collected in real world situations. Wild scene dataset includes different distances, various illumination conditions and complex reflectance. Thus, removing reflections is more difficult than in the controlled scenes dataset. SIR dataset includes \textit{454} paired data divided into: Postcard with \textit{199} images, SolidObjects with \textit{200} images and WildScenes with \textit{55} images. For comparison we use PSNR and SSIM \cite{ssim} metrics. PSNR value provides the numerical differences between two images while SSIM captures the structural differences between them. Higher values represents better results. 

\subsection{Ablation Study}
\textbf{Architecture analysis.}
To prove the benefit of the proposed model we conduct an ablation study. We enhance a classical residual network by introducing upscale, downscale layers, adding context modules like channel attention and dilated convolutions and making it deeper. Thus, we compare the proposed architecture with the baseline model set as a ResNet18 adapted for our use-case. We used as training set a fusion between real and synthetic data generated by our data generation model. The objective function and training parameters are kept the same for both models.
The performance comparison can be seen in the first two rows of Table~\ref{tab2}. Improved model obtain better results due to the added context modules. Channel attention mechanism exploit the inter-dependencies among feature channels while dilated convolutions allows to capture contextual information by progressive growing the receptive field of the network. Thus, global spatial information captured from the input image enables the context understanding of reflection and background layers that help to efficiently reconstruct the transmission image.

\begin{table}[t!]
\begin{center}
\caption{Ablation study.}
\resizebox{0.47\textwidth}{!}{
\begin{tabular}{|c|c|c|c|}
\hline
\textbf{ Architecture} & \textbf{Data generation} & \textbf{PSNR} & \textbf{SSIM} \\
 \hline
 ResNet & Baseline & 20.36 & 0.868\\
 \hline
 ReflectNet & Baseline & 22.80 & 0.877\\
 \hline
 ReflectNet & Proposed &\textbf{23.77} & \textbf{0.894}\\
 \hline
\end{tabular}}
\label{tab2}
\end{center}
\end{table}

\textbf{Data generation analysis.}
In this experiment we analyze the advantage of our proposed data generation model over the simple image blending model presented in Equation~\eqref{eq1}. The baseline data generation model is represented by an alpha-blending between a transmission and a reflection image. We use the collected images to generate two different training datasets, using the two methods. Then, the network is trained on each dataset for $90$ epochs. Results are presented in the last two rows of Table~\ref{tab2}. It can be easily seen that the proposed data generation model plays an important role in the performance of our method. Thus, it is important to generate a wide range of reflections to help the network to generalize on real data. 

\textbf{Computational complexity.}
In this paper, we have not only focused on the performance of the proposed model, but we have also taken into account the computational complexity regarding the running time and the size of the model. Most likely a reflection removal solution will take place in a desktop, mobile or web application which requires the model size to fall within certain limits and the execution time to be as small as possible. Thus, we carefully designed the model so that an efficient trade-off between performance and complexity is achieved compared with other state-of-the-art methods. 
We measure the execution time for processing one image of size $512 \times 512$ both on CPU and GPU devices. We used a system with an Intel Core i7 processor, \textit{2.6 GHz} with \textit{16GB} RAM memory and a system with a graphical processor NVIDIA GeForce RTX 2080 Ti with \textit{11GB} dedicated GPU memory. Results are shown in Table~\ref{tab3}.

\begin{table}[t!]
\begin{center}
\caption{Computational complexity.}
\resizebox{0.47\textwidth}{!}{
\begin{tabular}{|c|r|r|r|r|}
\hline
\textbf{Method}&\multicolumn{2}{|c|}{\textbf{Execution time}}& \textbf{Params} &\textbf{Model size}\\
\cline{2-3}
  & CPU & GPU &  &\\
 \hline
 BDN \cite{bdn} & \textbf{1.6s} & 0.012s & 75,183,994 & 300.70 MB \\
 \hline
 ERRNet \cite{errnet} & 42.6s & 0.015s & 86,693,125 & 330.00 MB\\
 \hline
 ReflectNet & 4.7s & \textbf{0.006s} & \textbf{2,291,635} & \textbf{8.79 MB}\\
 \hline
\end{tabular}}
\label{tab3}
\end{center}
\end{table}

% The following commented table contains additional column in GPU execution time. First column represents the avg time when the first run/execution was taken into account, while the second column represents the avg time when the first run/execution was NOT taken into account.

% \begin{table}[h]
% \begin{center}
% \resizebox{0.45\textwidth}{!}{
% \begin{tabular}{|c|r|c|r|r|}
% \hline
% \textbf{Method}&\multicolumn{2}{|c|}{\textbf{Execution time}}& \textbf{Params} &\textbf{Model size}\\
% \cline{2-3}
%   & CPU & GPU &  &\\
%  \hline
%  BDN & 1.6s & 0.025 / 0.012s & 75,183,994 & 300.70 MB \\
%  \hline
%  ERRNet & 42.6s & 0.10 / 0.015s & 86,693,125 & 330.00 MB\\
%  \hline
%  ReflectNet & 4.7s & 0.034 / 0.006s & 2,291,635 & 8.79 MB\\
%  \hline
% \end{tabular}}
% \caption{Computational complexity.}
% \label{tab3}
% \end{center}
% \end{table}

\begin{table*}[t!]
\begin{center}
\caption{Comparison between proposed method and state-of-the-art methods on SIR benchmark dataset. Higher values for PSNR and SSIM indicates better results.}
\resizebox{\textwidth}{!}{
\begin{tabular}{|p{2cm}|p{2cm}|p{2cm}|p{2cm}|p{2cm}|p{2cm}|}
\hline
\textbf{}&\multicolumn{5}{|c|}{\textbf{Methods (PSNR/SSIM)}} \\
\cline{2-6} 
\textbf{Dataset} & \textbf{\textit{CEIL} \cite{ceilnet}}& \textbf{\textit{PL} \cite{zhang2018single}}& \textbf{\textit{BDN} \cite{bdn}}  & \textbf{\textit{ERRNet} \cite{errnet}} & \textbf{\textit{Ours}} \\
\hline
Postcard& 19.98 / 0.800 &  15.80 / 0.597 & 20.54 / 0.849 & 21.84 / \textbf{0.886} & \textbf{22.00} / 0.880\\
\hline
SplidObject& 23.17 / 0.831 & 22.14 / 0.775 & 22.70 / 0.830 & 24.69 / 0.903 & \textbf{24.99 / 0.904}\\
\hline
WildScene& 20.84 / 0.794 & 21.15 / 0.828 & 22.00 / 0.825 & 23.80 / 0.857 & \textbf{25.75 / 0.907}\\
\hline
Average& 21.48 / 0.812 & 19,24 / 0.703 & 21.66 / 0.837 & 23.33 / 0.890 &\textbf{23.77 / 0.894}\\
\hline
\end{tabular}}
\label{tab1}
\end{center}
\end{table*}

\subsection{Comparison with State-of-the-art methods}
In this section we compare our network with state-of-the art methods for reflection removal like: CEIL \cite{ceilnet}, PL \cite{zhang2018single}, BDN \cite{bdn} and ERRNet \cite{errnet}.

Results are presented in Table~\ref{tab1}. It can be seen that our method produces better results in terms of PSNR and SSIM on all three subsets from SIR dataset. We noticed that existing methods perform poorly on Wild Scene subset because of their inability to generalize on real data. This happen mostly because there is no sufficiently large training dataset in research community and existing methods generates synthetic data for training. It can be seen that our method achieves best results on Wild dataset due to our data generation model that produces various and realistic reflection data for training. 

\begin{figure*}[t!]
  \centering
  \includegraphics[width=\textwidth]{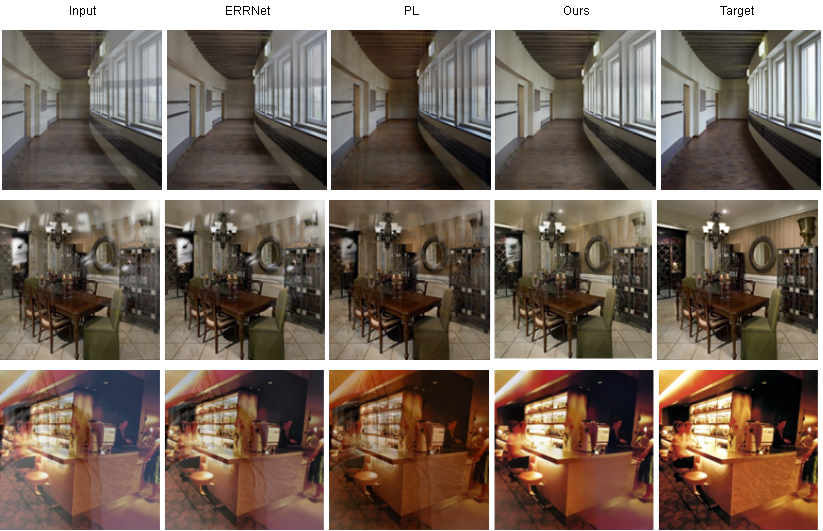}
  \caption{Comparison with state-of-the-art methods.}
  \label{fig:comparison}
\end{figure*}

Figure~\ref{fig:comparison} displays visual results on our generated data. It can be observed that both ERRNet and PL methods attenuate the reflection layer, but cannot completely eliminate it. Due to the included context modules, our method analyzes the global information in the input image and effectively suppresses the reflection layer. Also, it can be observed that our method produces realistic results with natural tones due to the adversarial training.

\section{Conclusion}

In this paper we proposed an automatic solution for reflection removal problem using deep learning techniques. Proposed architecture includes context modules that efficiently extract the underlying knowledge from training data which help to discriminate between reflection and background layers. We also addressed the problem of the lack of real paired training data which leads to poor performance of current deep learning-based reflection removal methods. We proposed a complex data generation model that produces realistic synthetic data covering a wide range of real-life reflections. Comparison of the SIR benchmark dataset shows that our approach outperforms state-of-the-art reflection removal methods both visually and objectively through PSNR and SSIM measures.

\bibliographystyle{IEEEtran}
\bibliography{references}

\pagebreak

\section{supplementary results}
\begin{figure*}[ht!]
  \centering
  \includegraphics[width=\textwidth]{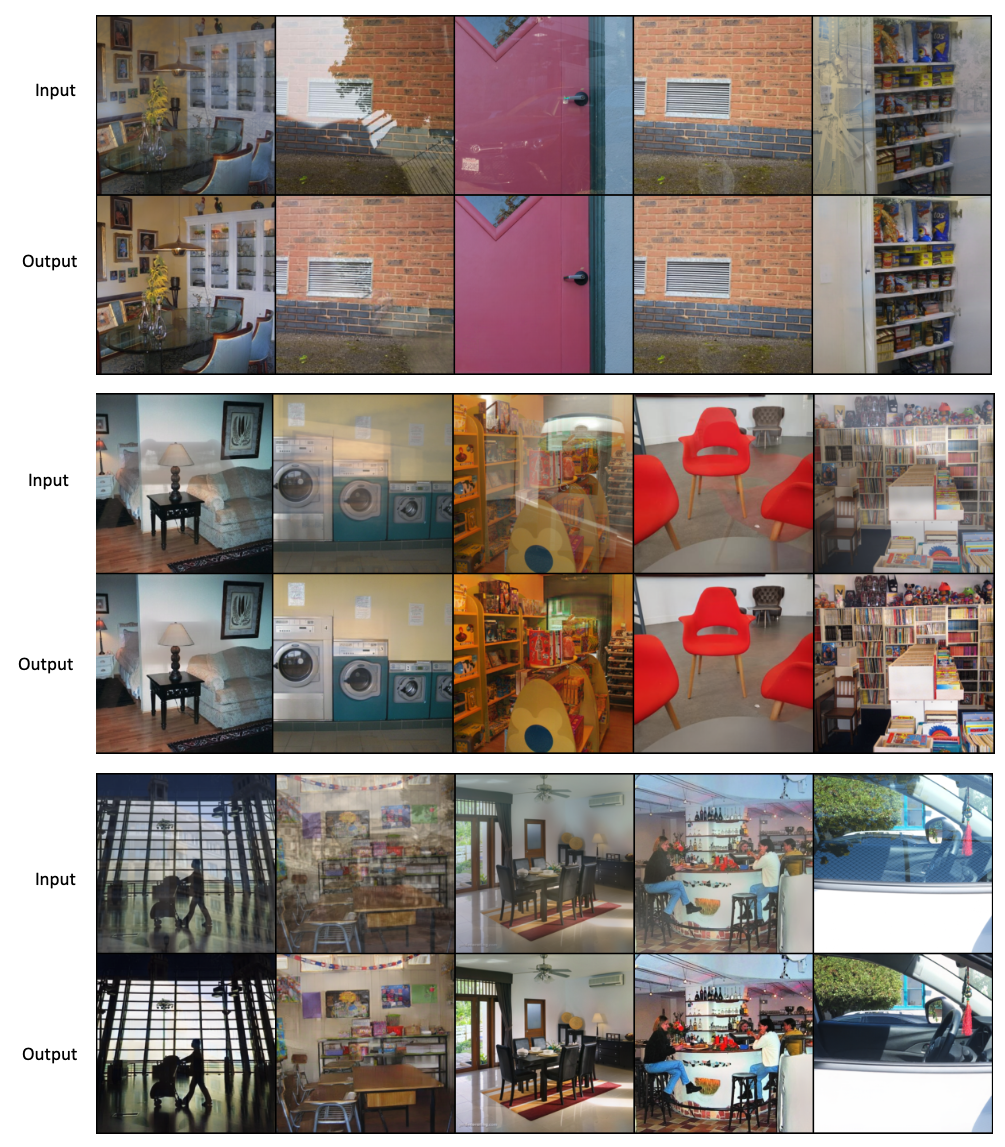}
  \caption{Results produced by our proposed method.}
  \label{fig:rez1}
\end{figure*}

\begin{figure*}[ht!]
  \centering
  \includegraphics[width=\textwidth]{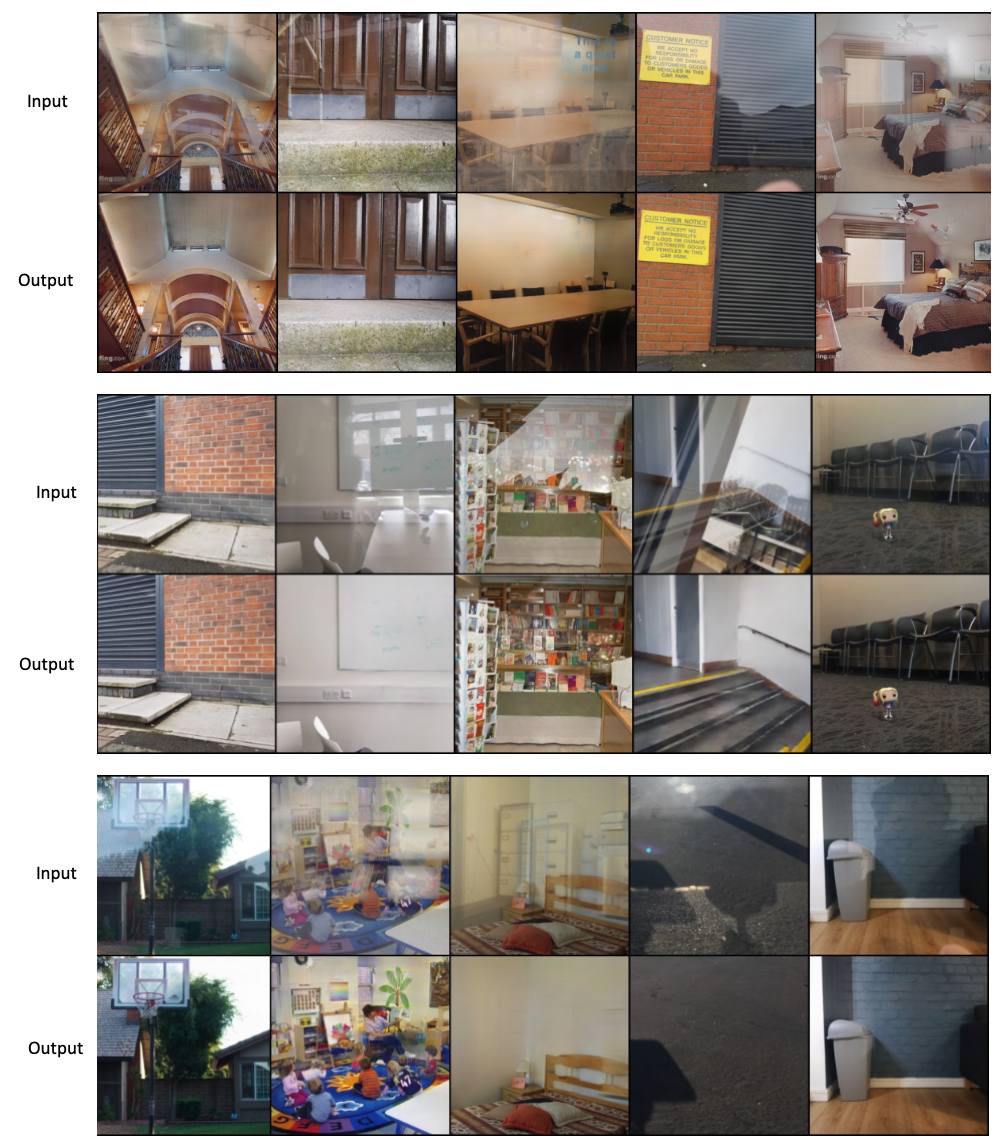}
  \caption{Results produced by our proposed method.}
  \label{fig:rez2}
\end{figure*}

\begin{figure*}[ht!]
  \centering
  \includegraphics[width=\textwidth]{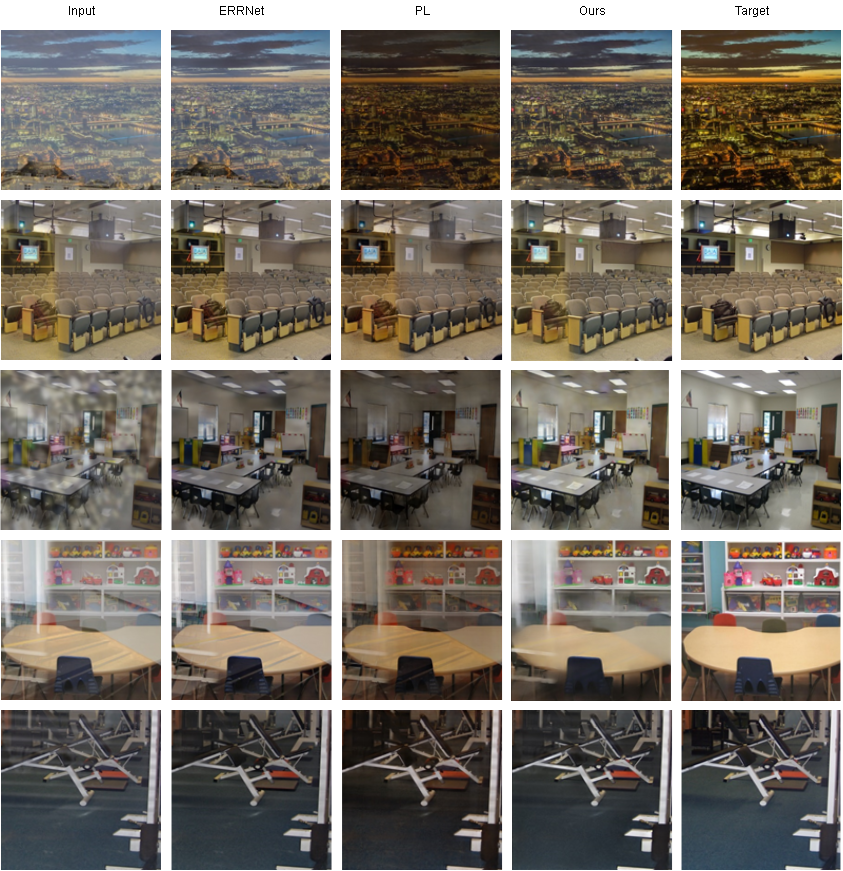}
  \caption{Comparison with State-of-the-art methods.}
  \label{fig:comp}
\end{figure*}

\begin{figure*}[ht!]
  \centering
  \includegraphics[width=\textwidth]{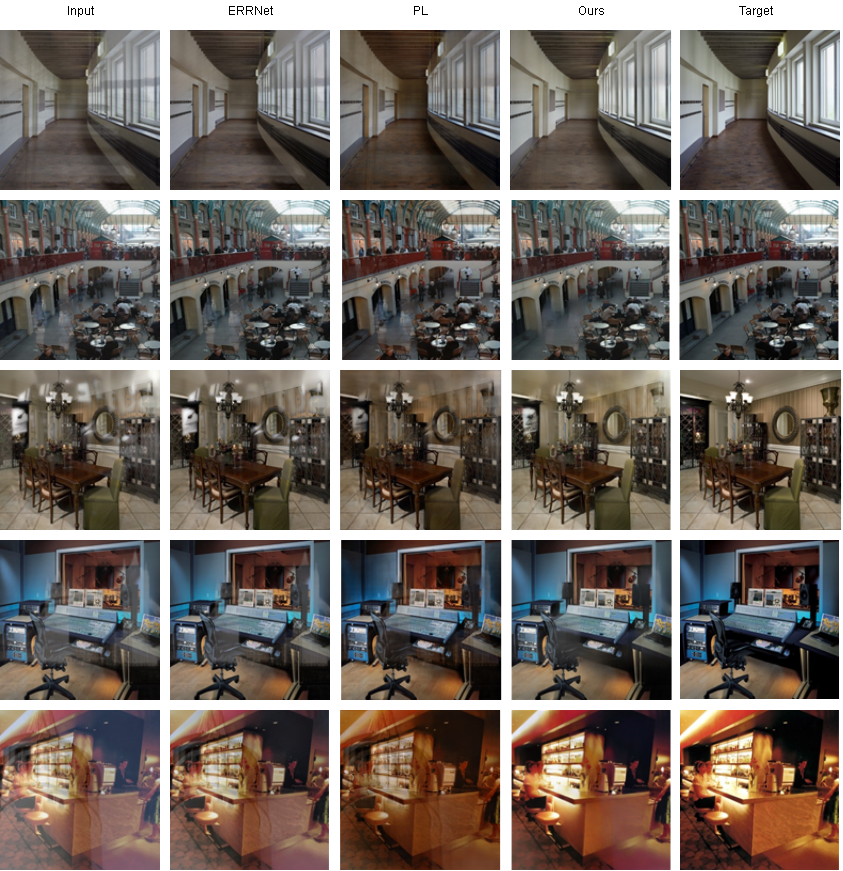}
  \caption{Comparison with State-of-the-art methods.}
  \label{fig:comp2}
\end{figure*}

\end{document}